# A Simulation-Based Method for Testing Collaborative Learning Scaffolds Using LLM-Based Multi-Agent Systems

Han Wu[a], Lishan Zhang[a]*, Chunming Lu[b]

[a]*School of Education, Beijing Institute of Technology, Beijing, People's Republic of China;* [b]*State Key Laboratory of Cognitive Neuroscience and Learning & IDG/McGovern Institute for Brain Research, Beijing Normal University, Beijing, People's Republic of China*

*Corresponding author: Lishan Zhang, Address: No. 5 South Zhongguancun Street, Haidian District, Beijing, 100081, P.R. China.

*Abstract*

**Background:** Traditional research on collaborative learning scaffolding is often time-consuming and resource-heavy, which hinders the rapid iteration and optimization of instructional strategies. LLM-based multi-agent systems have recently emerged as a powerful tool to simulate complex social interactions and provide a novel paradigm for educational research.

**Objectives:** This study proposes an LLM-based multi-agent simulation approach to investigate collaborative learning processes and the effectiveness of instructional scaffolds prior to actual classroom deployment. The research specifically examines the feasibility of simulating group discussions and the alignment of these simulations with established learning science theories.

**Methods:** The simulation system was implemented using the MetaGPT framework and GPT-4o, comprising one teacher agent and five distinct student roles (Leader, Supporter, Expounder, Rebutter, and Summarizer). Two scaffolding strategies—

"Deep Think before Speak" and "Direct Speak"—were compared across ten classical Chinese poetry appreciation tasks. Evaluation was conducted through discourse analysis of quality and behavior.

**Results and Conclusions:** The introduction of the "Deep Think before Speak" scaffold significantly improved the agents' discourse diversity and interaction depth while notably reducing content repetitiveness. Behavioral analysis showed that the scaffold encouraged more complex interaction patterns, such as reflecting, rebutting, and explaining. These findings align with the ICAP framework, as the scaffold prompted agents to move from simple "Active" participation to "Constructive" and "Interactive" knowledge co-construction. This study demonstrates the feasibility and ecological validity of using LLM-based multi-agent systems to simulate authentic collaborative learning dynamics.

**Keywords**

Large language model, Multi-agent systems, Collaborative learning, Generative AI, Instructional Scaffolds

# 1. Introduction

Collaborative learning is widely known as an effective method to improve learning efficiency. In such learning context, students as groups are expected to co-construct upon each other, so that every group member can get deep understandings towards the learning objects. However, it is not straightforward to come up efficient scaffolding to lead group co-constructions. It often requires extensive studies for different learning context. Traditionally, researchers raised some brilliant ideas (e.g. a way to scaffold a discussion for a group of students) and tested these ideas by running experiments with human participants, and hopefully get significant improvements in teaching or learning efficiencies. The researchers may improve the existing designs, adjust different experiment variables (e.g. different versions of one type of scaffolding), and run more rounds of experiments to get deeper understanding regarding the learning circumstance to be studied. This kind of studies may last several months and involve hundreds of human participants if everything gets smooth. It is very common that researchers find a scaffolding unsuccessful after running the actual experiments. We could not eliminate such possibilities for sure, but could we somehow predict the results given a designed scaffolding, so as to improve the successful rate? Large language models (LLMs) driven agent technology can provide an option to this question.

  Recently, researchers have begun to leverage the powerful generative and reasoning capabilities of LLMs to construct multi-agent systems that simulate complex and realistic social interaction processes (J. Zhang et al., 2025).

Sophisticated, model-driven narratives can effectively catalyze cooperative behaviors within digital societies, particularly when combined with strategic network positioning (De Curtò & De Zarzà, 2025). By leveraging prompt engineering and fine-tuning technologies, systems such as S3 enable agents to perceive environmental information and authentically mimic human behavioral patterns (Gao et al., 2023). Evaluations of these agent-based models show that the simulated group-level phenomena are highly consistent with empirical real-world data (Gurcan, 2024). However, LLMs and multi agent technologies are merely used in simulating interactions in the field of education. Moreover, as far as we know, no one has established simulated collaborative learning environments to explore the effectiveness of different types of instructional scaffolding, which is widely recognized as a critical mechanism for promoting deep collaborative learning (Verdú & Sanuy, 2014). By providing a structural framework, scaffolding helps learners progress from lower levels of engagement to higher cognitive levels (Kollar, Wecker, & Fischer, 2018). To promote learners' deep learning, various scaffolding strategies have been deployed to bridge the gap between concrete and abstract thinking (Pierce & Gilles, 2021). As we noted before, making effective scaffolding typically requires time-intensive Design-Based Research (DBR) or expert pedagogical design (Hsu, Lai, & Hsu, 2015). Therefore, we proposed a simulated collaborative learning system based on LLMs and multi-agent technologies, and evaluated the effectiveness of different types of scaffolding on collaborative learning in the simulated system.

Our simulated system includes one teacher agent and several student agents. The agents solve tasks by discussion just like human students in typical collaborative learning tasks. In specific, we implemented the simulated system based on MetaGPT framework, employed the prompt engineering to construct two types of metacognitive scaffolding, and assumed that the student agents strictly followed the taught scaffolding to conduct their discussion. The first type of scaffolding is "Deep Think before Speak" and the second type of scaffolding is "Direct Speak". The first type of scaffolding should lead more high-quality interactions than the second one, because existing studies showed that individual's internal self-explanation or knowledge construction (i.e. deep think) can increase interaction qualities during collaborative learning (Chi, De Leeuw, et al., 1994; Chi & Wylie, 2014; Ploetzner et al., 1999). By constructing the simulated collaborative learning systems and running experiments with the system, we aim to answer the following two research questions:

(1) How can the group discussion in collaborative learning be simulated based on LLMs and multi-agent technologies such as MetaGPT?

(2) Are the findings in the simulated collaborative learning system consistent with the existing qualitative studies with human participants?

The rest of the paper first briefly reviews the existing literature and theoretical foundations in related fields such as Computer assisted Collaborative Learning (CSCL), Multi-Agent Systems (MAS), and AI for Social Science. Then, we detail the design methods and evaluation framework for simulated collaborative learning experiments. Next, we report and analyze the results from the perspectives of discourse quality and discourse behavior. Finally, we conclude with remarks.

**2. Related work**

*2.1 Computer assisted collaborative learning*

Collaborative learning is defined as a situation in which two or more people attempt to learn together through coordinated, synchronous activities aimed at constructing and maintaining a shared conception of a problem (Dillenbourg, 1999; Roschelle & Teasley, 1995). During this process, group members work together through collaboration, in-depth discussion, and information exchange to complete learning tasks, share learning outcomes, and ultimately achieve common learning objectives (Miyake, 2007). The essence of collaborative learning lies in cooperation and communication, allowing students to deeply analyze and co-construct knowledge through dialogue, discussion, and debate. However, traditional collaborative learning models often face numerous challenges, such as lack of coordination in task division among group members, difficulty in synchronizing collaboration time, and some students avoiding or disengaging from discussions (Shea, 1995). To overcome these obstacles, Computer-Supported Collaborative Learning (CSCL) emerged, rapidly becoming a focal point for research and application among scholars. For instance, Järvelä, S. and Hadwin (2013) proposed the use of CSCL environments from the perspective of shared regulation, emphasizing the role of technology in supporting learning regulation within complex collaborative tasks. The impact of CSCL on collaborative learning and the broader field of education has been empirically tested.

Jeong et al. (2019) conducted a meta-analysis based on studies from 2005 to 2014 and found an overall effect size of 0.51 for CSCL in STEM fields. This effect size is considered moderate but highly significant in educational research, strongly supporting the overall advantages of CSCL in STEM education.

Within the recent years, AI agents based on large language models have emerged as a new medium to offer personalized and dynamic regulatory support in learning environments. This development has further expanded the research boundaries of traditional CSCL. Several studies have confirmed the significant advantages of AI agents in enhancing students' learning experience. For instance, Wang et al. (2025) developed an AI-agent to enhance students' programming learning. Compared to traditional CSCL environments, this agent-supported collaborative learning significantly improved students' self-efficacy and academic performance while effectively reducing cognitive load at the same time. AI agent also showed its effectiveness in second language learning, where students gained higher social and cognitive presence during interactions, and achieved better learning outcomes (X. Wang, Pang, Wallace, Wang, & Chen, 2024). Besides act as facilitators, AI agents can also act as teachable agent. Students can then improve their own knowledge and abilities by teaching the AI agent how to program (Chen, Wei, Le, & Zhang, 2024).

In summary, the research development of collaborative learning has demonstrated that the intelligent technology is becoming more and more important in this field. However, to the best of our knowledge, existing studies only considered using intelligent technology or AI agent to facilitate collaborative learning instead of simulating the process of collaborative learning. To address this research gap, this study used multi-agent technology to simulate collaborative learning process.

*2.2 Large Language Models and AI agent*

LLMs has provided key technological foundations for the development of AI agent technologies including both single-agent and multi-agent ones. With very few prompts, LLMs can enable AI agents to analyze complex data and provide contextually relevant responses (Örpek, Tural, & Destan, 2024). Agents driven by Multi-modal Large Models like GPT-4o can understand and generate different types of information like text, voice, image, video and etc (Islam & Moushi, 2025). An increasing number of studies have begun to apply these LLMs-driven agents in the field of education. These agents can interact with students, with other agents, or with both. Students can get learning benefits by interacting with the agents and observing the agents' interactions.

Based on the number of agents involved, applications in the educational domain can be divided into single-agent and multi-agent systems. Single-agent systems can be easily used to construct conversational based intelligent tutoring systems like AutoTuor (Graesser et al., 2004), which can guide students to engage in deep thinking through natural dialogues. Such systems used to require a lot of efforts in building the appropriate dialog scripts before the advance of LLMs. Recently, researchers are more focus on exploring the educational applications of multi-agent systems, which are capable of decomposing complex tasks into smaller and more manageable ones that are solved by corresponding dedicated agents (Jiang et al., 2024). For example,

Mohamedhen et al. (2024) built a multi-agent system based on convolutional neural networks and multilayer perceptrons, applying the Felder-Silverman model to first accurately identify students' learning styles, then provide personalized content recommendations. Viswanathan et al. (2022) designed several teaching agents with different functions for an online education system, forming an adaptive system capable of providing meaningful and targeted course content to learners. Zhang et al. (2025) developed a novel intelligent classroom framework, SimClass, and allowed real learners to interact with this simulated dynamic teaching environment across multiple courses, and showed that the multi-agent architecture significantly enhanced student engagement and learning outcomes.

It has no doubt that LLMs and AI agent technologies have been quickly applied in education after the notable achievement of GPT, and now researchers have started to build educational applications with multi-agent technology. In this study, we further explored how we could use the multi-agent technology to simulate collaborative learning.

*2.3 AI for social science*

The growing human-like reasoning and cognitive abilities of LLMs and AI agents have gained significant attention from social science researchers, prompting them to explore how this technology can be leveraged to adjust or even reshape research practices in the social sciences (Grossmann et al., 2023). Xu et al. (2024) assessed the relationship between AI and social science, highlighting the dual role of LLMs in research: as powerful research assistants and as reliable experimental agents simulating human behavior.

The application of LLMs as research assistants has been particularly notable, greatly enhancing research efficiency and the precision of data processing. With their robust natural language understanding capabilities, LLMs can automate the thematic analysis and coding of complex qualitative data (Zhang, Wu, Duan, & Du, 2025), significantly improving the efficiency of data analysis and inter-coder reliability. Bryda and Sadowski (2024) introduced a semi-supervised coding approach that applies AI algorithms to the coding and thematic analysis of free-text interviews, effectively enhancing the precision of qualitative data analysis and reasoning efficiency. Farjam et al. (2024) systematically guided researchers in using LLMs for automated coding in content analysis, emphasizing the numerous advantages of this approach in scalability, multilingual coverage, and cost-effectiveness. LLMs can also play a crucial role in generating research hypotheses and conducting systematic literature reviews (Bazgir & Zhang, 2025).

In terms of human behavior simulation, LLMs and AI agent technology are used to build platform that can run experiments for exploring and predicting human behaviors in the real world. Aher et al. (2023) argue that LLMs driven agents are the most reliable ones, capable of testing theoretical hypotheses under large-scale and reproducible conditions. For example, Park et al. (2022) designed a social simulation technique based on LLMs to model and generate responses from community users, as well as simulate social interactions. This approach helps social computing designers conduct better user behavior analysis or identify potential marginal situations that could lead to societal collapse. Horton (2023) analyzed three classic behavioral economics studies that used GPT-3.5 as the experimental subject. Their findings showed that the simulation experiments qualitatively reproduced conclusions similar to those from real human experiments, providing strong support for the feasibility of using agents for social science simulations. This simulation approach is low-cost, can accommodate arbitrary sample sizes, and allows for the testing of various prompt or parameter configurations, while also avoiding ethical concerns related to human participants. Guo (2023) enabled GPT to understand the rules of strategic game experiments and perform decision-making reasoning with human-like responses, uncovering potential patterns and logic observed in the games. Furthermore, Park et al. (2023) developed an interactive sandbox environment consisting of 25 agents, allowing users to interact with them and observe and evaluate the individual and emergent behaviors produced by the agents.

As Bail (2024) noted, this LLMs and AI agent-based simulation infrastructure is not only essential for ensuring broad access to high-quality research tools, but also vital for gaining a deeper understanding of the social forces that guide human behavior, particularly in the context of AI advancements. In this study, we built such simulation, and explored its usability in the context of collaborative learning.

## 3. Collaborative learning simulation design

*3.1 The simulated collaborative learning mechanism*

This study uses MetaGPT framework to construct the simulated collaborative learning system, integrating the DeepSeek-v3 model for prompt optimization, with GPT-4o serving as the core driving force to simulate collaborative learning scenarios.

The overall system architecture comprises three core components: the Communication Module, the Memory Module, and the Agent Module. These modules collaborate closely to establish a highly interactive and intelligent collaborative learning environment, with the system architecture diagram shown in Fig.1. During operation, the Communication Module manages the flow of information by employing a message routing mechanism, thus ensuring accurate information exchange between the teacher and student agents. The Memory Module stores and provides historical dialogues, which ensures that agents make their decisions based on the complete contextual information. Furthermore, the Agent Module enhances cognitive processing by setting Chain of Thought (CoT) and implements personalized behaviors through the role parameter settings.

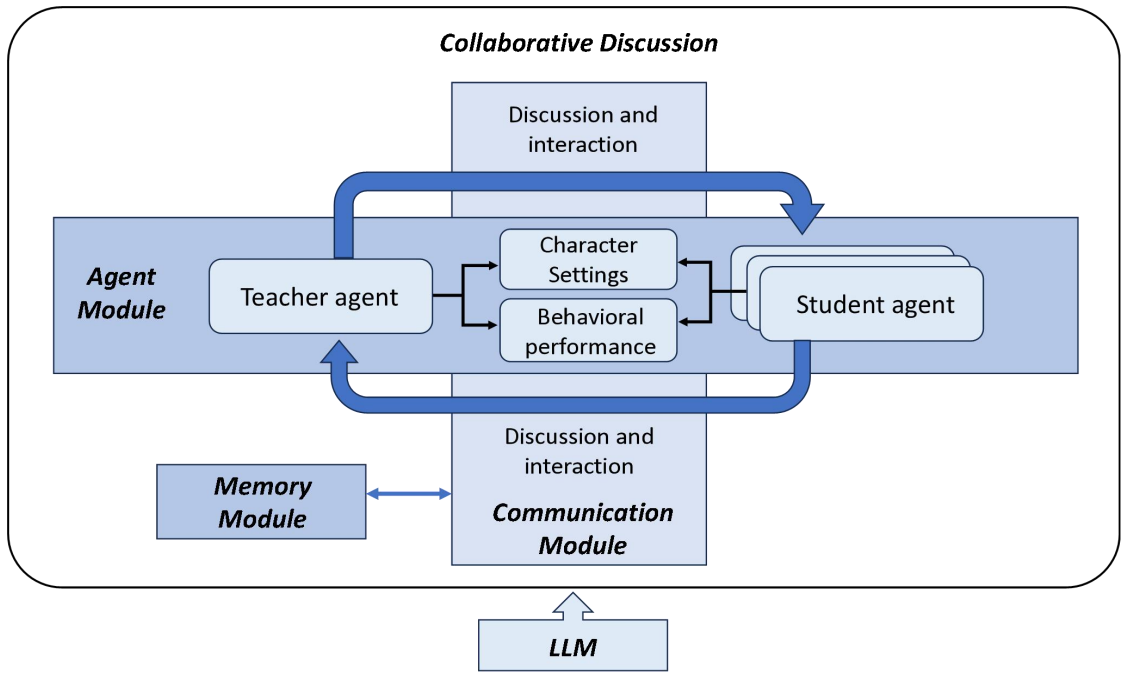

Fig.1 Architecture of the multi-agent collaborative learning system

The system operates in a task-driven cycle, guiding the discussion by focusing on predefined scoring points. In the initial phase, the teacher agent clarifies the collaborative task and key knowledge points, gives initial instructions, and sets the first round's speaking order of the student agents. Once receiving the instructions, each student agent reflects on the information based on their personal traits and previous interactions, and generates a response. After each discussion round, the teacher agent evaluates the coverage of knowledge points, updates the remaining knowledge points to be addressed, and provides feedback along with the next speaking order. This cycle repeats until all scoring knowledge points are covered. Additionally, student agent is designed to interact with each other based on collaborative learning practices. They can observe the contributions of other agents and adjust their behavior accordingly. For instance, when one agent offers a high-quality insight, other agents may choose to support, challenge, or further elaborate on the point, forming a natural discussion chain. The next section elaborates the detailed designs of the agents.

*3.2 The designs of the agents*

The system is designed with two types of agent roles: the teacher agent and the student agent, as shown in Fig.2 The teacher agent is responsible for commenting on students' discussions, controlling the pace of the conversation, imparting key knowledge points, and answering questions. Meanwhile, the student agents emulate learner behaviors, including posing questions, participating in interactive discussions, and absorbing knowledge. This differentiated role assignment endows the agents with human-like traits, significantly enhancing the interaction experience and realism in online collaborative learning contexts.

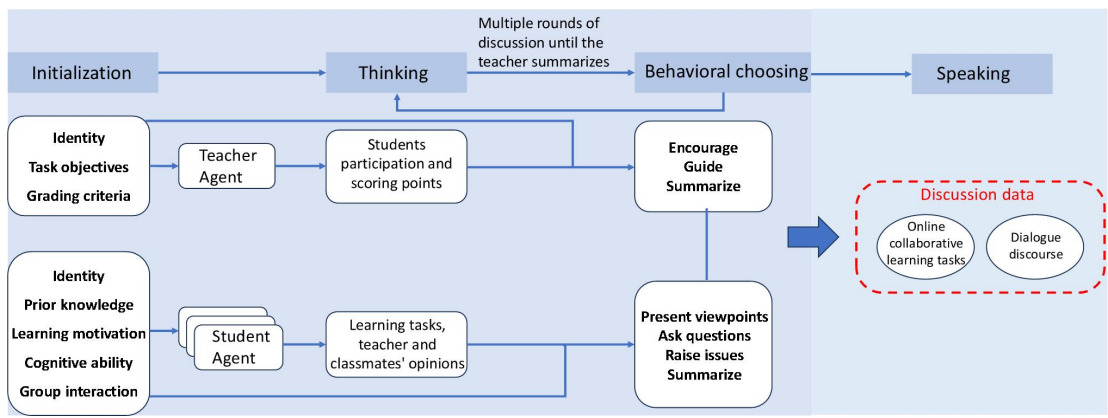

Fig.2 Design of teacher and student agent

The teacher agent is responsible for guiding and managing the collaborative learning process, ensuring the completion of tasks. Its main functions include: initialization, thinking, behavioral choosing, and speaking. The initialization function ensures that the teacher agent is clear about its identity, task objectives, and grading criteria. The data structure of its identity information is shown in Table 1. The thinking function allows the teacher agent to assess student participation and scoring points during discussions, so as to analyze the students' performance. The behavioral choosing function provides the teacher agent with three types of operational choices: 1. The ability to comment and ask questions based on the discussion in the previous round, and adjust speaking orders for the next around; 2. The ability to encourage or guide students to explore key points of the collaborative tasks; 3. The ability to provide feedback on each student's strengths and areas for improvement after multiple rounds of discussion. The speaking function integrates the above functional modules and generates teacher speech content with a word limit of 150 characters, taking the results of thinking as the guide, and executing the corresponding behavior selection.

Table 1. The identity information data structure of the teacher agent

| Field | Data type | Explanation |
| --- | --- | --- |
| id | int | Unique identifier of the teacher agent. |
| name | str | The name of the teacher agent, used in conversations. |
| base_definition | str | Including basic information such as professional field, teaching experience, etc. |
| learning_goal | str | Collaborative tasks and objectives scoring criteria. |

The student agent simulates the behaviors of real students and participates in the collaborative learning process. Its functions are also divided into four steps: initialization, thinking, behavior choosing, and speaking. The initialization function allows the student agents to clarify its identity information, prior knowledge, and specific role of the predefined behavior. The data structure of students' identity information is shown in Table 2. The thinking function enables the student agent to generate thoughts based on the tasks assigned by the teacher and the discussion history, and to make behavior choices by combining the group members' opinions and its own initialization parameters. The behavioral choosing function offers four operational choices: presenting viewpoints, questioning, raising issues, and summarizing: 1. Once the student agents have formed their own insights, they can proactively express views and share thoughts and understanding. 2. When they hear viewpoints that differ from their prior knowledge, they can raise questions to verify the accuracy and completeness of the information. 3. When they encounter content or topic which they want to discuss further, they can ask questions and actively seek knowledge. 4. At a certain stage of the discussion or at the end of the discussion, they also can summarize, review the key points and results of the discussion, and consolidate their learning outcomes. The speaking function is similar to that of the teacher agent. Based on the thinking and selected behavioral strategies, the student agents output speech content that conforms to their role settings, with a word count of about 80 words.

Table 2. The identity information data structure of the student agent

| Field | Data type | Explanation |
| --- | --- | --- |
| id | int | Unique identifier of the students' agent. |
| name | str | The name of the students' agent, used in conversations. |
| base_definition | str | Including basic information such as grade and major, etc. |
| assigned_role | str | The specific role for students' agent with predefined behaviors. |

## 4. Evaluation methods

### 4.1 Collaborative learning tasks

The collaborative learning tasks used in this study were derived from the Chinese poetry appreciation literacy course at a university in central China. These tasks were based on actual collaborative discussion tasks used in the course, which were then compiled into the collaborative learning task set for the appreciation of classical Chinese poetry in this study. To ensure the scientific validity and reliability of the assessment of the collaborative learning simulation, LLMs were used to generate five reference scoring criteria for each task. These criteria were reviewed and revised by the course instructors. In the end, a high-quality task set consisting of 10 collaborative learning tasks for classical Chinese poetry appreciation was formed (a full version is attached in the Appendix). Each task includes the original text of the poem, one collaborative learning task, and five scoring criteria. The example is illustrated in Table 3.

Table 3. Example of collaborative learning task set

| Poetry | Collaborative learning task | Scoring criteria |
|---|---|---|
| An overnight east wind blows, spring returns to the half-dead roots. Jade flowers respond to frost and snow, and the fruit Conceives the Universe. How could there be such beauty in the mountains and rivers, yet so majestic as to attract phoenixes and cranes? I only call upon the hermit to come here and climb this day. | Please answer the meaning of plum blossoms in this poem. | Tenacious life. "Spring Returns to Half-Dead Roots" depicts the resurgence of a dying plum blossom in the spring breeze, symbolizing the tenacity of the remnants of the Ming and Qing dynasties (Qu Dajun and others) in clinging to life amidst the tragedies of the Ming and Qing dynasties, and implicitly conveying the hope of national restoration. |
| | | Noble character. " Jade flowers respond to frost and snow " invokes the image of divine jade from the Chu Ci, imbuing the plum blossom with the characteristic of actively responding to suffering with its noble essence, breaking away from the traditional "passive, resistant to the cold" paradigm of plum blossom poetry. |
| | | The integrity of the remnants of the Ming and Qing dynasties. "The Fruit Conceives the Universe" borrows the image of a single fruit remaining in the "Peeling" hexagram from the Book of Changes, elevating the plum blossom's fruiting into a philosophical statement by the remnants of the Ming and Qing dynasties (Qu Dajun) to safeguard the flame of civilization in troubled times. |
| | | The spirit of seclusion. "How could there be such beauty in the mountains and rivers" negates the beauty of famous mountains and scenic spots, highlighting the spiritual heights of the small temple where the ancient plum blossom resides. " phoenixes and cranes" constructs a moral space for the remnants of the Ming and Qing dynasties that transcends the mundane. |
| | | Cultural inheritance. "Come here and climb this day" alludes to the metaphor of picking herbs in "Chu Ci", transforming the act of appreciating plum blossoms into a collective ritual for the remaining groups (such as the Lingnan School of Poetry) to pass on Chinese civilization through literature. |

To comprehensively evaluate the impact of different types of scaffolding on collaborative learning, this study utilized two approaches focusing on both the structural characteristics and the cognitive depth of student interactions. Specifically, assessment was conducted across two distinct dimensions: discourse behavior and discourse quality. Discourse behavior assessment was employed to quantify and categorize the frequency and patterns of student actions. Discourse quality assessment, in contrast, measured the epistemological and cognitive depth of the contributions to determine the intellectual value of the resulting dialogue.

*4.2 Assessment tools*

*4.2.1 Discourse quality coding framework*

The coding framework was developed by synthesizing and adapting quantitative metrics from recent studies on LLM evaluation to the specific context of simulated collaborative learning. Drawing on the work of Rai et al. (2024), who established standards for linguistic naturalness and syntactic correctness, we defined the fluency dimension to ensure the readability and grammatical precision of agent discourse. To make sure the logical consistency, we integrated Liu et al.'s (2024) metrics for internal logic with the contradiction detection methods proposed by Mündler et al. (2023). Furthermore, informed by Rossi et al.'s (2024) analysis of generative data challenges in social sciences, we incorporated the diversity dimension to explicitly measure the innovation and variability of viewpoints. The proposed framework for assessing the discourse quality is shown in Table 4. This framework includes five core dimensions: fluency, repetitiveness, contradiction, relevance, and diversity. Specifically, fluency focuses on the accuracy of grammar and the naturalness of expression; repetitiveness assesses the similarity between current utterances and previous ones; contradiction examines the consistency of an agent's viewpoints, identifying potential logical conflicts; relevance evaluates the alignment of content with task objectives; and diversity emphasizes the novelty of viewpoints and the breadth of knowledge. Since the teacher agent does not generate viewpoints, diversity is not used for coding its discourses. By analyzing these dimensions, the study not only evaluates the fluency of dialogues but also explores the logical consistency of agent behavior and the innovation in viewpoint expression.

Table 4. The framework of discourse quality coding

| Quality dimension | Coding rules | Judgment basis | Example |
| --- | --- | --- | --- |
| Fluency | Yes-1 <br> No-0 | Is the sentence grammatically correct? Is it in line with the daily expression habits of teachers and students? | "'The plum blossom blooms in response to the frost and snow, its abundant fruits bear fruit in the universe.' Can this be understood as the plum blossom symbolizing resilience and hope?" (High Fluency, 1) <br><br> "The plum blossom symbolizes resilience. It blooms in winter, so it's strong." (Disorganized Sentence, 0) |
| Repetitiveness | Yes-1 <br> No-0 | Are the viewpoints, arguments, expression logic and teacher interventions highly consistent with historical statements? | When discussing the symbolic meaning of plum blossoms, after the first mention of "plum blossoms symbolize perseverance and hope," each subsequent mention of the same argument and evidence will be counted as one repetition (repeating historical speech will be counted as one repetition). |
| Contradiction | Yes-1 <br> No-0 | Does the speech conflict with previous speeches, division of tasks or objective facts? | The student first proposed that "plum blossoms symbolize the hermit spirit," but then denied that "the hermit spirit is too idealistic" (self-contradiction, counted as 1 contradiction) |
| Relevance | Yes-1 <br> No-0 | Does the speech revolve around the poem and the task objectives, and does it make a substantial contribution to solving the problem? | Task Objective: "Please analyze the meaning of the lotus image in poetry." <br><br> Speech Content: "The lotus's 'icy and jade-like' appearance symbolizes purity." (Related) |
| Diversity | Yes-1 <br> No-0 | Are students' opinions innovative, multi-perspective, or do they cite cross-disciplinary knowledge? | "The plum blossom symbolizes an escape from reality and quotes a line from Tao Yuanming's poetry." (Highly innovative, counted as 1 highlight) |

Each dimension in the table is encoded independently using a binary system, with logical parallelism rather than exclusivity between the dimensions. This allows a single data point to be simultaneously marked as '1' across multiple dimensions. The corpus produced by the student and teacher agents for all tasks in each experiment was collected. Each speech was then evaluated for quality along five or four dimensions, and the number of samples for each dimension was accumulated.

*4.2.2 Discourse behavior coding framework*

In collaborative learning research, discourse behaviors analysis serves as an effective evaluation tool, playing a crucial role in understanding the interaction patterns among learners. The coding framework for the discourse behaviors of student agents was developed by synthesizing established categories from human collaborative learning research and tailoring them to the communicative characteristics of LLMs. We derived the irrelevant learning behaviors dimension from Adams et al. (2002), specifically incorporating their observations on "watching passively" and "disengaged" behaviors. To capture the social dynamic of the interaction, we referred to the taxonomy of Tan et al. (2022), which emphasizes "learning outcomes" and "social interactions and processes" as essential application indicators of AI in collaborative learning. Furthermore, we adapted from the verb-dominated coding scheme proposed by Wang et al. (2020), which focuses on studying behavioral patterns in online collaborative learning environments with different learning material formats. The coding framework for student agent's discourse behaviors is shown in Table 5. It includes four key dimensions: irrelevant learning behaviors, social relational behaviors, collaborative analytical behaviors, and viewpoint construction behaviors.

Table 5. The framework of discourse behavior coding

| Behavioral Dimension | Behavior Type | Coding Rules | Example | Coding |
|---|---|---|---|---|
| irrelevant learning behaviors | Ineffective | Do not speak or are irrelevant to the discussion. | Do not speak or are irrelevant to the task objective. | A1 |
| social relational behaviors | Plan | The discussion begins by bringing out topics related to the topic and task under discussion and proposing directions. | Let's start the discussion with... What are everyone's thoughts on this aspect? | B1 |
| | Monitor | During the discussion, monitor the completion of task objectives based on historical speeches and supplement the discussion with new directions. | We had a very thorough/inspiring discussion. We covered..., and we can move on to... | B2 |
| collaborative analytical behaviors | Reflect | A statement summarizing the results of the discussions so far. | We covered... from... to..., covering a wide range of topics. | C1 |
| viewpoint construction behaviors | Elaborate | Speech that constructs one's own opinions through citations, examples, etc., or constructs one's own cognitive system based on the speeches of others. | I think..., because... | D1 |
| | Support | Start by clearly stating your agreement with the other person's point of view and use examples to illustrate it. | I agree with..., because... | D2 |
| | Question | During the discussion, raise questions about points that you do not understand. | I have some questions about what you meant by... | D3 |
| | Rebut | During the discussion, question opinions that you disagree with. | I think your statement is incorrect/one-sided, because... | D4 |
| | Explain | Answer and explain other people's questions. | Regarding the issue of..., I think/feel... | D5 |

Because the teacher agent just guides the discussion instead of really involving it, only three types of codes were used for coding teacher agent's discourses: Encouragement (A1), Guidance (B1), and Summarization (C1). Each type of code corresponds to a distinct teaching goal. Encouragement (A1) fosters students' motivation and participation through positive feedback, especially when students offer valuable insights or engage actively in discussions. Guidance (B1) involves directing the focus of students through questioning, prompts, or topic redirection to help them concentrate on learning objectives and explore key concepts. Summarization (C1) typically occurs at the end of a discussion, aiming to consolidate and synthesize students' viewpoints, assisting in the formation of a systematic knowledge framework and reinforcing learning outcomes.

*4.2.3 LLM assisted assessment tool*

Conduct of traditional discourse analysis and deductive coding needs considerable time and labor involved. The coding process can also bring researchers' subjective biases. Therefore, we also explored the feasibility of applying LLMs to automatic deductive coding tasks (L. Zhang et al., 2025). Drawing on the work of Tai et al. (2024), who applied LLMs in text data analysis, this research further investigates the specific potential of LLMs in discourse analysis. In the initial phase of the study, two researchers pre-coded 20% of the data samples to ensure the reliability and validity of the evaluation framework used. Subsequently, GPT-4o was employed as an auxiliary coding tool to provide automated coding support for evaluating discourse quality and behavioral performance during the collaborative process. After completing the automatic deductive coding, a manual sampling verification is conducted. 20% of the already coded data is randomly selected again to calculate the consistency between the model's coding and the manual coding. This approach not only enhances the efficiency of data analysis but also provides reliable methodological support for subsequent collaborative learning research.

 The prompt design for GPT-4o's automatic deductive coding consists of two core components: (1) a detailed coding specification explanation, which defines the criteria, judgment standards, and boundary conditions for each coding category, and includes representative examples as well as edge cases; and (2) a mandatory self-explanation requirement, in which the model is instructed to provide explicit reasoning for its coding decisions. The purpose of this design is to enhance the interpretability of the coding results while also facilitating subsequent verification and necessary interventions by researchers. The complete prompt content can be found in Appendix.

## 5. Experiment design

Remind that student agents in the simulated collaborative learning system were assumed to strictly follow the taught scaffolding (i.e. "deep think before speak" or "direct speak"). We used prompt engineering to define how the student agents process information and act differently in the two conditions. Specifically, a student agent in the "direct speak" condition generated utterances in direct response to other student/teacher agents' inputs, without engaging in additional processing mediated by explicit prompts or workflows. On the other hand, a student agent in the "deep think before speak" condition generated utterances guided by a deep cognitive process described with prompt engineering. The cognitive process mainly encompasses the four steps below:

- Content analysis: Accurately understanding the content and theme of the poem;
- Instruction interpretation: Comprehending the teacher's instructions and the expected direction of the discussion;
- Context tracking: Paying close attention to the key points in the speech of other student agents;
- Differentiated contribution: Formulating unique insights or problems to address.

The design of the scaffold's prompts is shown in Table 6., and the complete prompt design can be found in the appendix. The specific experimental process is shown in Fig. 3.

Table 6. Design of the scaffold's prompts

| Stage | Prompt |
|---|---|
| Think | {<br>  "Background": "You are in a poetry appreciation class and need to brainstorm based on an assignment given by the teacher. There is one teacher and five students. You need to each play a role in the discussion and work together to complete the group task."<br>  "Poetry": "{{poem}}",<br>  "Dialogue History": "{{context}}",<br>  "Teacher's Instructions": "{{latest_instruction}}",<br>  "Reflection Guidelines": "Please reflect on the current discussion and the teacher's instructions based on your role. You need to:<br>  1. Understand the content and theme of the poem<br>  2. Understand the teacher's instructions and expectations<br>  3. Pay attention to the key points of other students' statements<br>  4. Consider what unique insights or questions you can offer."<br>  "Output Template":<br>  {<br>    "Understanding of the Poem": "Based on my role, my understanding of this poem is...",<br>    "Reaction to Others' Comments": "My thoughts on other students' comments are...",<br>    "Possible Contributions": "Considering my role, the unique perspective or insight I can offer in the discussion is...",<br>    "Inner Thoughts": "As {{name}}, my true thoughts at this moment are..."<br>  }<br>} |

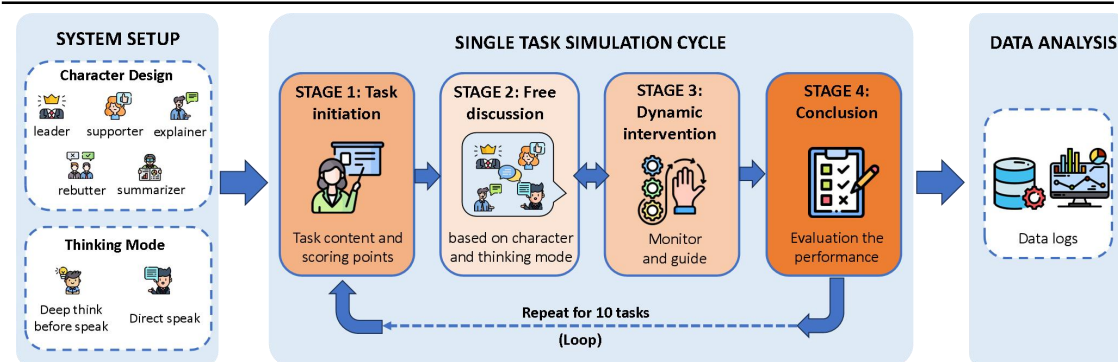

Fig. 3. Experimental workflow of collaborative learning simulation

Each student agent was assigned to a specific role based on the role collaboration theory (Benne & Sheats, 1948). This theory classifies the roles that emerge during group interactions into three broad categories: Group Task Roles (e.g., Expounder, Opinion Giver), Group Building and Maintenance Roles (e.g., Encourager, Coordinator), and Individual Roles (e.g., Dominator, Help Seeker). Drawing upon this theoretical framework and considering the specific needs of collaborative learning, we identify and define five distinct student agent roles: Leader, Supporter, Expounder, Refuter, and Summarizer. These five roles encompass key functions ranging from task progression to relationship maintenance, aiming to simulate a diverse range of collaborative interactions. The detailed descriptions of all the five roles are shown in the Table 7. All the dialog data generated by the agents are recorded for the discourse analysis afterward, so that we can compare how the agents in the two conditions perform differently.

Table 7. Five types of student agent roles

| Role design | Role function | Expected performance of the role |
| --- | --- | --- |
| Leader | Start a topic and maintain a discussion | Raise the core issues for discussion, guide the direction of the dialogue, ensure that the discussion does not deviate from the topic, propose new perspectives when the discussion stagnates, and promote the participation of all members |
| Supporter | Citation, explanation, and answering questions | Provide theoretical basis or textual evidence to support viewpoints, explain complex concepts in depth, clarify other members' confusion, and expand the depth and breadth of discussions |
| Explainer | Choose a point of view and support it with examples | Identify and reinforce valuable insights, enhance the persuasiveness of arguments through specific examples or analogies, create a positive discussion atmosphere, and encourage members to continue thinking deeply |
| Rebutter | Ask questions actively and question critically | Challenge weaknesses or holes in existing perspectives, present counterexamples or alternative explanations, prompt the team to re-examine assumptions, prevent groupthink, and promote more comprehensive analysis |
| Summarizer | Summarize key points and reach consensus | Regularly summarize the key points discussed, integrate different viewpoints, clarify the interim results of the discussion, help the team form a common understanding, and point out unresolved issues |

The experiment included ten Chinese poetry appreciation tasks. Each task is divided into four main phases. The first phase is the task initiation, during which the teacher agent clearly defines the collaborative task content and the knowledge points involved, establishing basic speaking norms to lay the foundation for the subsequent in-depth discussions. The second phase is free discussion. The student agents generate utterances based on their roles and cognitive process logic. In each turn, a student agent may choose actions such as offering opinions, raising objections, or remaining silent. The student agents usually engage in multiple rounds of interactions to explore various issues of the given task. The number of rounds for solving a task depends on when all the knowledge points of the task are covered. In the dynamic intervention phase, the teacher agent focuses on monitoring two core indicators: the activation progress of knowledge points and the balance of student agents' participation. This phase is conducted concurrently with the second phase, with student agents engaging in discussions while teacher agent provides dynamic monitoring. Finally, in the conclusion phase, when the preset termination conditions are met, the teacher agent will comprehensively evaluate the performance of each student agent throughout the discussion process, assessing their 'cognitive progress level.' The teacher will also summarize, organize, and analyze all the discussed issues, providing appropriate 'comments and encouragement' to help the agents better understand and master the learned knowledge.

## 6. Results

*6.1 Descriptive analysis of agent discourse*

We first calculated the word counts and the ratio of utterances between teacher agent and student agent. The results are depicted in Fig. 4. The size of the bubbles represents the frequency of utterances. The more counts, the bigger the bubble, and it is evident that the discourse frequency of student agents significantly exceeds that of teacher agents, which aligns with our experiment setting. The data indicates that in terms of the proportion of discourse frequencies between teachers and students, the teacher-student ratio in the "think before you speak" mode is 1:3.9, which is better than the 1:3.6 ratio in the "speak directly" mode. This suggests that the former provides students with greater opportunities for autonomous expression. At the same time, the visualized results reveal that in the "deep think before speak" mode, the line between teachers and students are longer, with a greater disparity in the "depth of speaking" between them. This suggests that the mode provides students with more space for autonomous exploration, while reducing the frequency of immediate, high-density teacher interventions.

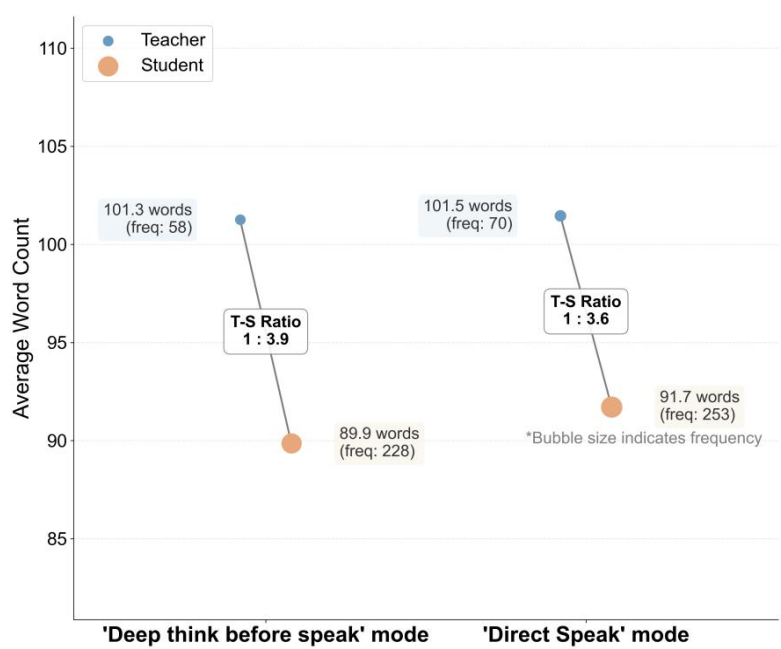

Fig. 4. Comparison chart of discourse text data of teacher and student agents

We also used an independent samples t-test to examine the impact of two modes on the length of discourse generated by the agents, as shown in Table 8. The results indicated that there was no statistically significant difference between the average discourse length in the "deep think before speak" mode (M = 124.5, SD = 22.3) and the "direct speak" mode (M = 98.2, SD = 18.7) (t = -1.239, p = .216). Additionally, the effect size (Cohen's d = 0.113) was well below the 0.2 threshold, further confirming that the difference in output volume between the two modes was negligible. This finding suggests that the addition of cognitive intervention scaffolding did not significantly alter the output scale of the agents. The reason for this may lie in the underlying constraints of the LLM, where the maximum length limit of the prompts and the predefined output style lead to relatively stable output lengths across different scaffolding strategies. From a research validity perspective, this consistency reflects that the pacing and information density of the interactions remained highly synchronized across both modes, thus validating the robustness of the multi-agent system in simulating collaborative learning under different intervention strategies.

Table 8. Comparison of utterance length between two modes

| Mode | N | mean | sd | *t* | *P* | Cohen's d |
|---|---|---|---|---|---|---|
| Deep think before speak | 253 | 91.72 | 17.57 | -1.2391 | 0.216 | 0.1132 |
| Direct speak | 228 | 89.86 | 15.00 | | | |

Overall, these characteristics of the discourse text data reveal that the "deep think before speak" mode contributes to building a more efficient and autonomous collaborative discussion environment, reducing ineffective interactions, and providing a solid data foundation for subsequent discourse quality and behavioral analysis.

*6.2 Qualitative analysis of agent discourse quality*

This experiment primarily focuses on a comparative analysis of the "deep think before speak" and "direct speak" cognitive modes, using discourse quality as the evaluation metric. Initially, we randomly selected 20% of the data as the pre-coding sample, which included 100 student utterances and 25 teacher utterances. After manual coding by two researchers, the Kappa value reached 0.75, demonstrating good internal consistency of the coding process. Subsequently, based on a consensus between the two researchers on the final coding results, we used a large language model to re-code the same dataset to adjust and verify the accuracy of the model's coding. The results showed a Kappa coefficient of 0.80 between the manual and model-based coding, further confirming the high consistency of the evaluation system and validating the model's ability to perform automated deductive coding. Finally, the large language model was used to automate the coding of the remaining data, with a final manual review and verification conducted after the coding process was completed.

*6.2.1 Between-subjects analysis of student agents*

The study conducted an independent samples t-test on the discourse quality coding data from ten tasks in each of the two groups, with the results detailed shown in Table 9. To control for the False Discovery Rate (FDR) across multiple dimensions, the Benjamini-Hochberg (BH) procedure (Thissen, Steinberg, & Kuang, 2002) was used to adjust the p-values. The analysis revealed that the "deep think before speak" mode significantly outperformed the "direct speak" mode in terms of the Repetitiveness (adjusted p =.012) and Diversity (adjusted p =.016). There were no statistically significant differences in Fluency (adjusted p =.340) or Relevance (adjusted p =.340), and the zero values for the Contradiction indicator further confirmed the technical stability of the LLM in maintaining grammatical norms and content coherence.

Table 9. The discourse quality of student agents under two thinking modes

| Discourse quality dimension | "deep think before speak" mode | "direct speak" mode | t | P | Adj. *p*(BH) |
|---|---|---|---|---|---|
| Fluency | 22.700±5.250 | 25.300±6.549 | -0.979 | 0.340 | 0.340 |
| Repetitiveness | 9.900±3.071 | 16.300±5.165 | -3.368 | 0.003 | 0.012 |
| Contradiction | 0.000 | 0.000 | N/A | N/A | N/A |
| Relevance | 22.700±5.250 | 25.300±6.549 | -0.979 | 0.340 | 0.340 |
| Diversity | 13.500±3.749 | 9.100±2.846 | 2.956 | 0.008 | 0.016 |

*Note: Adjusted p-values were calculated using the Benjamini-Hochberg procedure to account for the FDR.*

Focusing on the analysis of the differentiation indicator, the "deep think before speak" mode demonstrated a clear advantage. The average value for Repetitiveness decreased by 6.40, while Diversity increased by 4.40. This suggests that the multi-step iterative thinking mechanism can effectively deepen the cognitive processing of the agent. Its positive effects are mainly reflected in three aspects: first, through the guidance of the thought chain, it successfully reduced the redundancy of the content, thus improving the repetitiveness indicator; it activated a broader range of knowledge associations, thereby enhancing the diversity of the discourse; while improving the quality, it maintained the internal consistency of the semantic system, preventing the risk of generating false information (hallucinations). From a cognitive perspective, this framework is more aligned with human's deeply thinking processes, fostering the generation of more innovative, logically structured, and information-rich language expressions.

*6.2.2 Within-subjects analysis of student agents*

To further analyze the discourse quality differences among various collaborative roles in the "deep think before speak" mode, the study examined the distinctions in repetitiveness and diversity indicators for each role and illustrated their percentage distributions in Fig. 5. It's important to note that because the encoding of each dimension is logically parallel and not mutually exclusive, a piece of data can be encoded as either repetitiveness or diversity. For example, in the collected data, a student agent's statement is, "I strongly agree with Li Si and Wang Mei's views! Furthermore, I think 'The mountain air is beautiful at sunset, and the birds fly back to their nests' also perfectly embodies the meaning of seclusion. The tranquility of the mountains and the scene of birds returning to their nests echo the reclusive life symbolized by chrysanthemums, showcasing Tao Yuanming's transcendence and leisure." The first part of this statement supports and repeats the previous student's statement, so repetitiveness is encoded as 1. The second part, where the student proposes a combination of natural phenomena and the poet's personality, is an innovative viewpoint, so diversity can also be encoded as 1. The data analysis indicates that the discourse characteristics of each role are highly aligned with its predefined functional positioning.

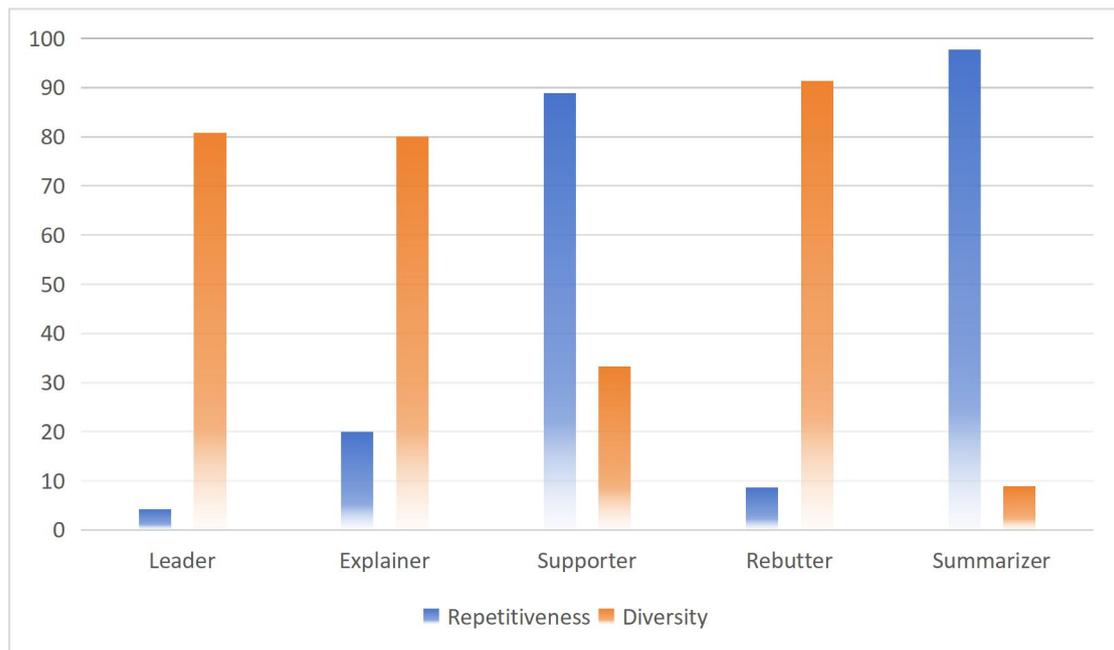

Fig. 5. The discourse quality of student agents in the "deep think before speak" mode

As the initiator and facilitator of the discussion, the leader exhibited the highest level of innovation and the lowest redundancy, with a diversity score of 80.85% and the minimum repetitiveness. This aligns precisely with the leader's core responsibility of initiating topics and maintaining the depth of the discussion. The diversity level of the explainer was similar to that of the leader, suggesting that the viewpoints expressed by the explainer were similarly novel, fitting the role's function of offering personal insights or building upon others' perspectives. The contrast between the high repetitiveness rate (88.89%) and low diversity of supporters reflects the dual characteristics of their content focus and convergent argumentation when reinforcing specific viewpoints. The rebutter, with its critical thinking, demonstrated the highest diversity while maintaining low repetitiveness, highlighting the role's dynamic advantage in challenging existing arguments and promoting divergent thinking. Finally, the summarizer presented an extreme distribution of discourse quality: a high repetitiveness of 97.78% and a low diversity of only 8.89%, visually reflecting the mechanism of building consensus through frequent restatement.

This role-based variation, driven by preset behaviors, effectively validates that the "deep think before speak" mode activates the functional attributes of each role. By balancing repetitiveness and innovation, this mode enables the construction of a diversified collaborative discussion ecosystem while ensuring content relevance. More importantly, it facilitates stronger functional differentiation in the quality indicators of different roles, providing robust theoretical and empirical support for the design and implementation of efficient collaborative learning models.

*6.2.3 Between-subjects analysis of teacher agents*

The study analyzed the performance of the teacher agent under both modes, as shown in Table 10, and found no statistically significant differences across any dimensions after applying the Benjamini-Hochberg (BH) procedure. While the Relevance indicator showed a nominal difference in the unadjusted test ($p = .042$), it did not remain significant after the BH correction. This observed trend in Relevance is likely due to the fact that in the "deep think before speak" mode, the teacher agent makes decisions on whether to intervene and guide based on the historical dialogue records. If the system determines that no intervention is needed, the teacher agent may provide encouragement or affirmation, in which case the content of its speech is not directly related to the collaborative task. In contrast, the teacher agent in the "direct speak" mode does not engage in such a reflective or judgmental process, and therefore, the content of its speech is more closely tied to the task itself.

Table 10. The discourse quality of teacher agent under two thinking modes

| Discourse quality dimension | "deep think before speak" mode | "direct speak" mode | t | P | Adj. p(BH) |
|---|---|---|---|---|---|
| Fluency | 5.800±1.135 | 7.000±1.885 | -1.724 | 0.102 | 0.102 |
| Repetitiveness | 1.400±1.074 | 2.200±0.919 | -1.789 | 0.090 | 0.102 |
| Contradiction | 0.000 | 0.000 | N/A | N/A | N/A |
| Relevance | 5.200±1.316 | 6.900±2.079 | -2.185 | 0.042 | 0.102 |

*Note: Adjusted p-values were calculated using the Benjamini-Hochberg procedure to account for the FDR.*

### 6.3 Qualitative analysis of agent discourse behavior

In terms of the discourse behavior coding results, the study also selected 20% of the total data for evaluation. The inter-rater reliability test of the manual coding (Kappa = 0.73, >0.60) demonstrated good consistency. Furthermore, the Kappa coefficient between the manually agreed coding results and the large model's coding results increased to 0.83, which strongly supports the high consistency of the evaluation system.

#### 6.3.1 Between-subjects analysis of student agents

A between-group comparative analysis of the discourse behavior distribution in student agents revealed distinct patterns, as detailed in Table 11. To control for the inflation of Type I error across these indicators, a Bonferroni-corrected alpha level of 0.0063 (0.05/8) was also applied. Under this adjustment, the "deep think before speak" mode demonstrated statistically significant differences in several key categories. Notably, a significant decrease was maintained in Elaborate (D1) (adjusted p = .008), alongside significant increases in Plan (B1) (adjusted p = .035), Explain (D5) (adjusted p = .035), Rebut (D4) (adjusted p = .044), and Reflect (C1) (adjusted p = .045). The BH-adjusted results confirm that the "deep think before speak" mode effectively encourages agents to adopt more diverse interaction strategies, thereby broadening and deepening the discussion.

Table 11. The discourse behavior of student agent under two thinking modes

| Discourse behavior dimension | "deep think before speak" mode | "direct speak" mode | t | P | Adj. p(BH) |
| --- | --- | --- | --- | --- | --- |
| Ineffective (A1) | 0.000 | 0.000 | N/A | N/A | N/A |
| Plan (B1) | 2.200±1.398 | 0.800±0.788 | 2.757 | 0.013 | 0.035 |
| Monitor (B2) | 2.600±1.578 | 3.200±1.932 | -0.761 | 0.457 | 0.457 |
| Reflect (C1) | 3.100±1.449 | 1.700±1.159 | 2.385 | 0.028 | 0.045 |
| Elaborate (D1) | 3.900±2.234 | 11.300±4.218 | -4.903 | 0.001 | 0.008 |
| Support (D2) | 4.700±2.057 | 5.500±1.649 | -0.959 | 0.350 | 0.400 |
| Question (D3) | 3.400±2.171 | 2.200±1.751 | 1.361 | 0.190 | 0.253 |
| Rebut (D4) | 1.700±1.418 | 0.500±0.527 | 2.508 | 0.022 | 0.044 |
| Explain (D5) | 1.200±1.229 | 0.100±0.316 | 2.741 | 0.013 | 0.035 |

*Note: Adjusted p-values were calculated using the Benjamini-Hochberg procedure to account for the FDR.*

To gain a deeper understanding of how student agent behaviors are transferred between the "deep think before speak" and "direct speak" modes, this study generated a behavior transition matrix through computational analysis, as shown in Fig. 6. These two heatmaps illustrate the transition probabilities between student agent behaviors under different thinking modes, where darker colors indicate higher transition probabilities. The vertical axis represents preceding behaviors, while the horizontal axis represents subsequent behaviors.

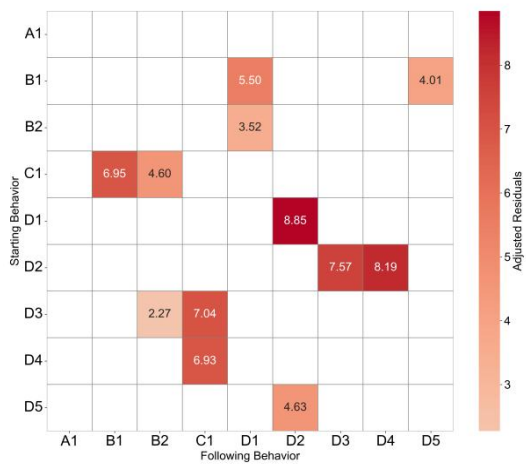

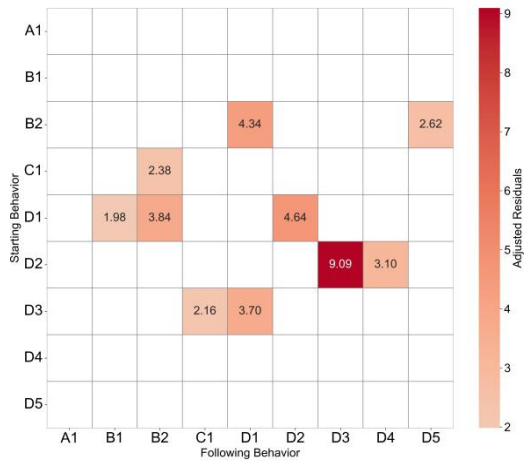

(a) "deep think before speak" mode          (b) "direct speak" mode

Fig. 6. Heat map of student agent's behavior transfer under two thinking modes

From the left heatmap in Fig 6.3, it is evident that the "deep think before speak" mode exhibits a diverse and highly interconnected behavior transition pattern. Notable transitions include: the shift from D1 to D2, which shows that articulating a viewpoint often triggers supportive behaviors, thus forming a positive feedback loop of collaborative knowledge construction; the high-probability transitions from D2 to D3 and D2 to D4 reveal that supportive behaviors can either trigger further questions or lead to rebuttals, highlighting the multidirectionality of cognitive development; the high transition probabilities from C1 to B1 and C1 to B2 demonstrate a natural flow from reflection and summary to the planning and monitoring of the next round, completing a full discussion loop; finally, the transitions from D3 to C1 and D4 to C1 suggest that questioning and rebuttal behaviors often lead to reflection, aiding deeper cognitive integration. These complex transition sequences collectively form a highly interactive behavioral network, covering the entire collaborative process from Plan (B1), through multi-faceted knowledge construction (Category D), to Reflect (C1).

In contrast, the right side of Figure 6.3 reveals that the "direct speak" mode exhibits a significantly simplified behavior transition pattern. The most notable feature is the extremely high transition probability from D2 to D3, which indicates that supportive behaviors are mostly followed by questions. However, these questions often lack subsequent in-depth follow-up, making it difficult to form a complete cognitive conflict resolution chain. A few other significant transitions include: D1 to D2, which reflects a linear pattern of support following the articulation of a point; B2 to D1 and B2 to D5, which show that behavior monitoring can still lead to knowledge expression; and D1 to B1 and D1 to B2, suggesting that after the articulation behavior, there may be a return to planning and monitoring, though the strength of these transitions is notably weaker than in the "deep think before speak" mode. In the "direct speak" mode, transitions related to Rebut (D4) and Explain (D5) are almost entirely absent, and the lower half of the behavior matrix shows large areas of empty space, indicating that deep critical interactions are severely limited.

By comparing the behavior transition matrices of the two modes, we find that the "deep think before speak" mode's behavior transition network is more complex and diverse, with more effective transition points and a more balanced distribution. In contrast, the "direct speak" mode's transition network is simpler and more linear, concentrating primarily on a few limited paths. This comparison strongly demonstrates that the "deep think before speak" mode not only helps balance the distribution of individual behavior types but also facilitates the construction of a more natural and complete behavioral flow network, allowing the entire discussion process to exhibit the dynamic evolution characteristic of real collaborative environments.

*6.3.2 Within-subjects analysis of student agents*

To more clearly illustrate the distribution characteristics of discourse behaviors across different roles in the "deep think before speak" mode, this study constructed a Role-Behavior Proportion Analysis Chart (as shown in Fig. 7.) and conducted an in-depth analysis. The chart explicitly reveals the differentiated distribution patterns of five collaborative roles across nine categories of discourse behaviors, reflecting a distinct functional division of labor.

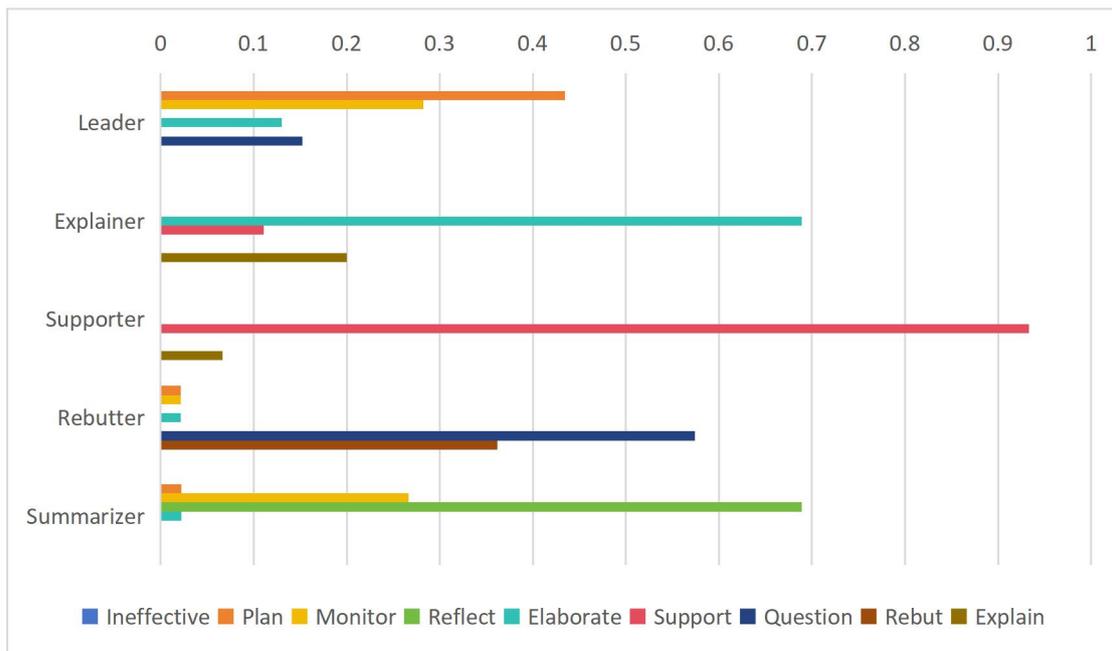

Fig. 7. The discourse behavior of student agents in the "deep think before speak" mode

The Leader role is characterized by a high proportion of planning behaviors (B1) at 43.48%, which helps establish the framework for the discussion, along with 19.57% of monitoring behaviors (B2) to adjust the discussion flow in real time. Additionally, the Leader role displays 44.74% of explanation behaviors (D1), effectively guiding the discussion toward the intended topic, fulfilling its core function of steering the conversation. The Explainer role leads in both explanation and elaboration behaviors, with the proportion of explanation behaviors being three to five times higher than that of other roles, aligning perfectly with its function of introducing novel ideas and offering in-depth explanations. The Supporter role shows a significantly higher proportion of supporting behaviors compared to all other roles, strongly emphasizing its function in reinforcing valuable insights. The Critic role, on the other hand, exhibits the highest proportions of questioning and rebuttal behaviors, highlighting its critical and interrogative nature. The Summarizer role has the highest proportion of reflection behaviors, consistent with its function of synthesizing key points and promoting consensus. Additionally, the Summarizer also demonstrates a relatively high proportion of monitoring behaviors (B2), indicating that this role actively introduces new discussion angles during reflective phases.

*6.3.3 Between-subjects analysis of teacher agents*

The behavioral differences in the teacher role between the two thinking modes are also noteworthy. As shown in Table 12, after applying the BH procedure, statistically significant differences were observed in both Guidance (B1) (adjusted $p = .015$) and Encouragement (A1) (adjusted $p = .030$). Specifically, the teacher agent in the "deep think before speak" mode demonstrated a higher frequency of encouragement and a lower frequency of direct guidance. These results suggest that when student agents engage in discussions using this structured thinking mode, the teacher agent can reduce the frequency of direct interventions and instead focus on providing encouragement and affirmation, thereby effectively promoting students' autonomous exploration and deeper cognitive engagement.

Table 12. The discourse behavior of teacher agent under two thinking modes

| Discourse behavior dimension | "deep think before speak" mode | "direct speak" mode | t | P | Adj. p(BH) |
|---|---|---|---|---|---|
| Encouragement (A1) | 2.100±0.567 | 1.100±1.100 | 2.554 | 0.020 | 0.030 |
| Guidance (B1) | 3.400±1.173 | 5.500±1.715 | -3.194 | 0.005 | 0.015 |
| Summarization (C1) | 0.300±0.674 | 0.400±0.516 | -0.372 | 0.714 | 0.714 |

*Note: Adjusted p-values were calculated using the Benjamini-Hochberg procedure to account for the FDR.*

Based on the analysis of both discourse quality and discourse behavior, the "deep think before speak" mode outperforms the "direct speak" mode in several core indicators: (1) Content novelty: There is a significant increase in discourse diversity and a notable decrease in redundancy; (2) Role alignment: The behaviors of each agent are more consistent with their predefined functional roles; (3) Interaction diversity: The proportion of behaviors reflecting deep interactions, such as support, rebuttal, and explanation, has increased; (4) Teacher intervention optimization: The teacher's encouragement behaviors have increased, while guidance behaviors have decreased, further enhancing students' autonomy. These findings provide important theoretical guidance for the future design of multi-agent collaborative learning systems: by embedding structured thinking processes within the agents, the quality and efficiency of their interactions can be significantly improved, making the simulated collaborative learning process more realistically replicate real classroom situations.

# 7. Discussion

This study constructs a multi-agent collaborative learning system based on the MetaGPT framework and designs comparative experiments to investigate the system's performance in simulating group discussion in collaborative learning. In this section, we will address the research questions based on the findings.

**RQ1: Feasibility and Characteristics of LLM-based Multi-Agent Simulation.**

The research findings substantiate the feasibility and effectiveness of the proposed multi-agent system as a novel platform for educational experiments. Data analysis reveals that the system not only successfully replicates the complex and dynamic interactive environment of real-world collaborative learning, but more importantly, the student agents demonstrate remarkable consistency with their characters. Whether as the leader responsible for framing the discussion or the challenger stimulating depth through critical thinking, the agents' discourse behaviors align closely with their functional roles.

The introduction of the "deep think before speak" mode proved to be a determining factor in this success. By simulating the metacognitive process of human learners, this mode enabled agents to exhibit a more structured and systematic pattern of behavioral transitions. Unlike the "direct speak" mode of chat interactions, the agents formed a self-sustaining discussion loop capable of effectively generating and resolving cognitive conflicts.

From a methodological perspective, this finding addresses the call within computational social science to utilize agents for simulating human social interactions, suggesting that LLM-based agents possess sufficient cognitive abilities to internalize complex social roles. Compared to traditional empirical research, which is time-consuming and difficult to control for variables, this high-fidelity and controllable simulation environment offers an efficient alternative for exploring collaborative learning mechanisms, enabling researchers to repeatedly test educational hypotheses at a very low cost.

**RQ2: Alignment between Simulated Behaviors and Human Collaborative Learning Patterns.**

Comparative analysis reveals a significant positive effect of the cognitive intervention scaffolding. In the dimension of discourse quality, the experimental group equipped with the cognitive scaffolding (the "deep think before speak" mode) exhibited higher content diversity and significantly reduced redundancy in their dialogues. This suggests that the mandatory "thinking" scaffold effectively suppresses various negative behaviors commonly observed in student collaboration, encouraging them to activate a broader knowledge network before speaking, thus generating more information-dense and innovative ideas.

This mechanism-driven enhancement of cognitive depth aligns with the ICAP framework (M. T. H. Chi & R. Wylie, 2014), which classifies students' learning activities based on their level of cognitive involvement, ranging from high to low, into four categories: Interactive, Constructive, Active, and Passive. Based on the ICAP classification criteria, the "deep think before speak" cognitive intervention scaffold designed in this study can promote students' agents to shift from simple "Active" to

"Constructive" and "Interactive" knowledge co-construction. This thinking phase simulates the knowledge construction process of real students, corresponding to the "Constructive" dimension in the ICAP framework. Meanwhile, the discussions, support, and explanations generated based on this process represent a higher level of "Interactive" participation.

In contrast, agents in the "direct speak" mode predominantly remained at the "Active" level. The fact that the simulated agents exhibited differentiated cognitive behaviors consistent with the ICAP hierarchy validates the system's fidelity. It demonstrates that LLM-based agents, like human learners, require explicit metacognitive scaffolding to transcend superficial participation and achieve deep knowledge co-construction.

Further analysis of collaborative performance reveals that the "deep think before speak" mode not only optimized individual outputs but also reshaped the group's interaction ecology. The study observed a significant increase in the proportion of deeper interactive behaviors, such as explanation and rebuttal, forming a more complex and interconnected network of behavioral transitions. This indicates that student agents no longer merely state their views in isolation, but instead engage in more frequent debates and negotiations to co-construct knowledge, resulting in a shift from "shallow interaction" to "deep collaboration." It is noteworthy that this increased student agency also prompted adaptive changes in the teacher agent's behavior—shifting from frequent direct instructions to more emotional support and encouragement. This phenomenon, where student autonomy increases and teacher intervention decreases, validates the positive impact of the cognitive intervention scaffolding on the quality of discourse and depth of discussion in collaborative learning. It further supports the ecological validity of the simulation system in replicating authentic educational dynamics.

## 8. Conclusion

This study is grounded in the theoretical foundations of collaborative learning, aiming to explore the application value and potential of the LLM-based multi-agent system in simulating collaborative learning. In specific, we first presented how to design and implement the simulation system based on the multi-agent framework named MetaGPT, and evaluated different scaffolding (i.e. deep think before speak v.s. direct speak) effectiveness on the simulation system. The comprehensive analysis showed that the results of the simulation system generally aligned with the learning science theories such as self-explain and ICAP. The detailed results also reported the possible fine-grained differences of collaborative learning discourses, which provided valuable references in practice.

While our study validates the theoretical efficacy of metacognitive scaffolding in the simulated collaborative learning system, future work is needed to explore how the acquired knowledge from the simulation system can guide the teaching and learning practice with human subjects.

**Ethics approval and consent to participate**

Not applicable. This study does not involve human participants or animals, and therefore ethical approval and informed consent are not required.

**Funding**

This work was supported by [BLINDED]. The funding body had no role in the design of the study, collection, analysis, and interpretation of data, or in writing the manuscript.